\documentclass[sigconf]{acmart}
\copyrightyear{2026}
\acmYear{2026}
\setcopyright{cc}
\setcctype{by}

\acmConference[MM '26]
  {Proceedings of the 34th ACM International Conference on Multimedia}
  {November 10--14, 2026}
  {Rio de Janeiro, Brazil}

\acmBooktitle{Proceedings of the 34th ACM International Conference on Multimedia
  (MM '26), November 10--14, 2026, Rio de Janeiro, Brazil}

\acmISBN{979-8-4007-2213-4/2026/11}
\acmDOI{10.1145/3767308.3835654}

\AtBeginDocument{%
  }

\usepackage{multirow}

\begin{document}

\title{HyperCD: Effective Source Detection in Hypergraphs via Causal Disentanglement}



\author{Yan Kang}
\affiliation{%
  \institution{Northwestern Polytechnical University}
  \city{Xi'an}
  \country{China}}
\email{kyan@mail.nwpu.edu.cn}

\author{Peican Zhu}
\authornote{Corresponding authors.}
\affiliation{%
  \institution{Northwestern Polytechnical University}
  \city{Xi'an}
  \country{China}
}
\email{ericcan@nwpu.edu.cn}

\author{Keke Tang}
\affiliation{%
  \institution{Guangzhou University}
  \city{Guangzhou}
  \country{China}
}
\email{kktang@mail.ustc.edu.cn}

\author{Le Cheng}
\affiliation{%
  \institution{Northwestern Polytechnical University}
  \city{Xi'an}
  \country{China}
}
\email{chengle@mail.nwpu.edu.cn}

\author{Zun Liu}
\affiliation{%
  \institution{Northwestern Polytechnical University}
  \city{Xi'an}
  \country{China}
}
\email{liuzun@nwpu.edu.cn}

\author{Ang Dong}
\affiliation{%
  \institution{Northwestern Polytechnical University}
  \city{Xi'an}
  \country{China}
}
\email{angdong@mail.nwpu.edu.cn}

\author{Yangming Guo}
\authornotemark[1]
\affiliation{%
  \institution{Northwestern Polytechnical University}
  \city{Xi'an}
  \country{China}
}
\email{yangming\_g@nwpu.edu.cn}

\renewcommand{\shortauthors}{Kang et al.}

\begin{abstract}
  Source detection in hypergraphs plays an important role in mitigating the spread of harmful rumors and ensuring social stability. Current methods frequently overlook the causal entanglement problem in observations, thereby limiting the performance of source inference. In this work, we propose a source detection approach in \textbf{Hyper}graphs via \textbf{C}ausal \textbf{D}isentanglement (HyperCD). To decouple causal propagation semantics from confounding noise, we propose the causal disentanglement module. A dual-branch variational encoder is utilized to disentangle snapshot embeddings into causal and confounding latent spaces, ensuring that the inference module relies solely on causal propagation signals. To further enforce this separation, an adversarial inference loss term is introduced to guarantee statistical independence between the decoupled branches. Moreover, to capture diffusion dynamics in hypergraphs, an attention-based hypergraph aggregation module is designed, which comprises the node-to-hyperedge propagation and hyperedge-to-node propagation processes. Extensive experiments on six datasets demonstrate the superiority of our HyperCD in effectively mitigating the causal entanglement problem.
\end{abstract}




\maketitle



\section{Introduction}
Driven by the rapid development of online platforms, social networks have become the primary channels for information dissemination and the exchange of opinions. However, this high degree of interconnectivity also facilitates the viral spread of harmful content, such as rumors and malicious propaganda \cite{DBLP:conf/mm/Zhang00ZLH024}, posing severe threats to public trust and social stability \cite{shelke2019source}. Consequently, source detection, which aims to identify the original spreaders of information cascades, has emerged as a critical issue in network science \cite{DBLP:conf/mm/ZhaoNCB23}. As hypergraphs provide a more faithful characterization of high-order group interactions in social networks, hypergraph-based source detection has emerged as a critical research direction.


\begin{figure}[htbp]
\centering
\includegraphics[width=0.48\textwidth]{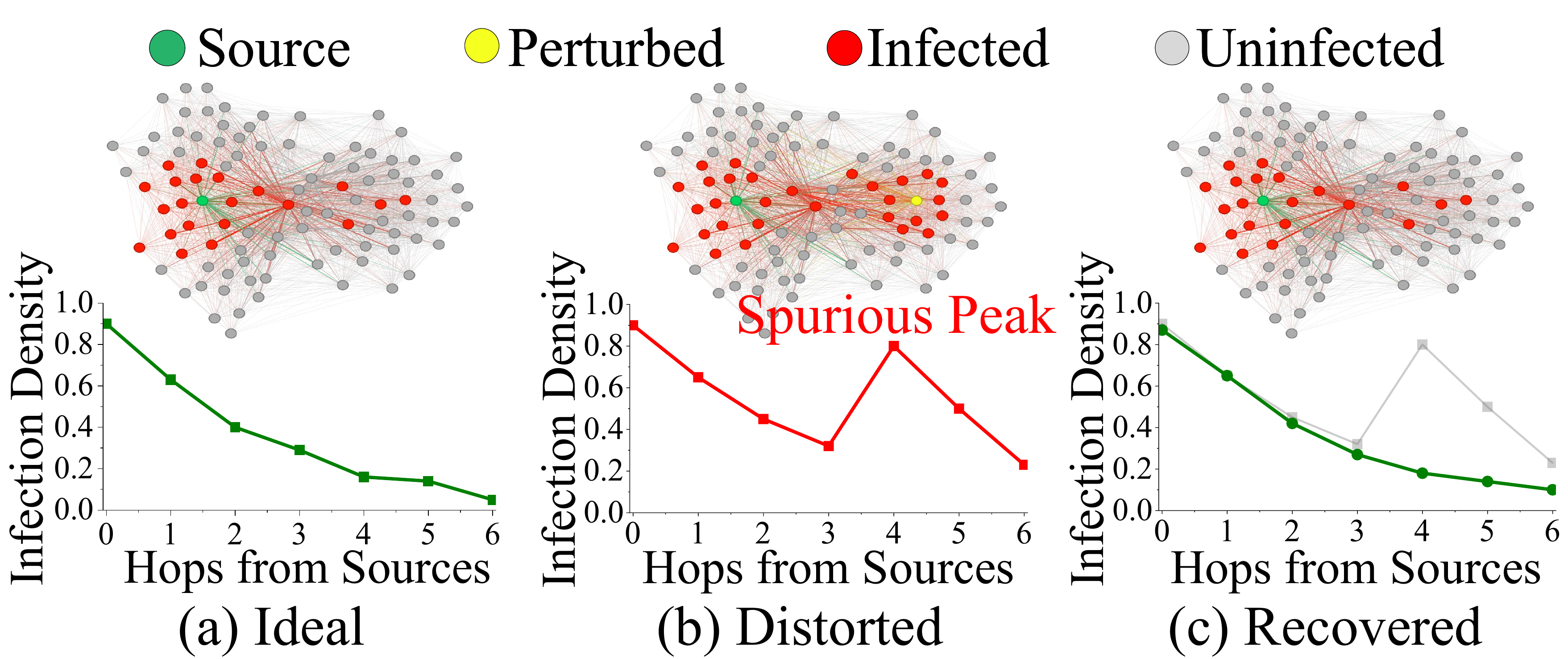}
\caption{The signal-noise entanglement problem in source detection. (a) Ideal infection density distribution of a propagation process. (b) Distorted infection density distribution by confounding environmental noise. (c) Recovered infection density distribution by performing proposed HyperCD, highlighting its effective disentanglement capabilities.}
\label{intro}
\end{figure}

Current source localization approaches for hypergraphs typically infer the source by learning diffusion dynamics from observed snapshots \cite{cheng2025hyperdet, cheng2025sourcedetmamba}. However, these methods implicitly assume a strictly causal relationship between the snapshots and the source, which fundamentally overlooks the issue of causal entanglement. In practice, the source detection process is susceptible to signal–noise entanglement, which introduces confounding effects into the inferred propagation patterns. As in Figure \ref{intro}, such confounders lead to a spatial shift of the propagation center and distort the distance-decay trend, thereby obscuring the true causal signals and degrading the performance of source localization. Furthermore, the causal entanglement problem is further pronounced in hypergraphs due to nonlinear collective influence and complex propagation patterns in higher-order interactions \cite{majhi2022dynamics}. Therefore, it is desirable to focus on causally relevant semantics while mitigating the influence of irrelevant noise semantics for source detection. 

The primary challenges of source detection in hypergraphs are summarized as follows: 1) \textbf{The causal entanglement problem in snapshots.} Most existing approaches infer sources by learning correlations from observed snapshots, implicitly assuming a strictly causal relationship. In practice, however, diffusion processes are inevitably perturbed by confounding noise, leading to the causal entanglement problem. Consequently, the performance of existing methods is significantly compromised in noisy, real-world scenarios. 2) \textbf{Complex diffusion dynamics in hypergraphs.} Due to the presence of high-order interactions, diffusion processes in hypergraphs are inherently more complicated than those in pairwise graphs. This complexity makes it challenging to accurately capture the underlying propagation dynamics from discrete snapshots, thereby limiting effective source detection.

Therefore, we propose an effective source detection approach in \textbf{Hyper}graphs via \textbf{C}ausal \textbf{D}isentanglement (HyperCD), consisting of a causal disentanglement module and an attention-based hypergraph aggregation module. Based on the observed snapshots, the causal disentanglement module decouples the causal propagation semantics from confounding noise via a dual-branch variational encoder architecture. Specifically, the snapshot embeddings are projected into two latent spaces corresponding to causal semantics and confounding semantics, ensuring that the inference module solely depends on the causal propagation semantics. To guarantee the validity of the disentanglement, the training loss integrates a mutual information minimization term and an adversarial inference term, enforcing that each branch effectively captures its respective semantics. Moreover, we propose an attention-based hypergraph aggregation module to capture high-order interactions. The hypergraph aggregation alternates between node-to-hyperedge and hyperedge-to-node propagation, where an efficient linear attention mechanism adaptively integrates messages from incident nodes and hyperedges.

The primary contributions are summarized as follows:

\begin{itemize}
\item To address causal entanglement in diffusion, we propose a disentanglement module that decouples propagation semantics from latent noise. This ensures that the inference process solely
depends on features relevant to diffusion, thereby enhancing source localization accuracy.

\item To capture complex diffusion patterns in hypergraphs, we develop an attention-based hypergraph aggregation module. By alternating between the node-to-hyperedge and hyperedge-to-node processes, this module efficiently extracts high-order dynamics from snapshot embeddings.

\item
Empirical results demonstrate that HyperCD significantly outperforms state-of-the-art baselines across widely utilized datasets, effectively mitigating the causal entanglement problem and capturing the diffusion dynamics inherent.
\end{itemize}

\section{Related Work}

\subsection{Source Detection in Pairwise Graphs}
Early source detection methods primarily rely on centrality-based heuristics, including degree centrality \cite{albert2000error}, betweenness centrality \cite{louni2014two}, and closeness centrality \cite{freeman2002centrality}. Centrality-based methods primarily rely on network topology and often ignore the underlying propagation dynamics. Subsequently, observer-based approaches incorporated limited propagation information from a restricted set of monitored nodes \cite{gajewski2022comparison, 11543370}. For example, Cheng \textit{et al.} integrated Monte Carlo-based sensor deployment with Bayesian inference for source detection in incomplete networks \cite{cheng2024sdsi}, while Hu \textit{et al.} introduced an improved LPSI algorithm designed for scenarios with limited observers \cite{hu2024source}. However, their performance strongly depends on observer deployment or specific diffusion models, which largely limit their practical applicability. 

Due to the convenience and feasibility of snapshot acquisition, snapshot-based approaches have been widely adopted for source detection. Wang \textit{et al.} introduced the LPSI approach, which applies a label propagation strategy guided by source prominence to identify diffusion origins \cite{wang2017multiple} and relies on specific diffusion patterns. Additionally, Dong \textit{et al.} proposed GCNSI \cite{dong2019multiple}, a GCN framework capable of locating multiple rumor sources without assuming a specific propagation model, whereas HFSD \cite{cheng2024heuristic}, IVGD \cite{wang2022invertible}, and SL-VAE \cite{ling2022source} incorporate dynamic propagation features prior to source inference. Furthermore, Cheng \textit{et al.} proposed GIN-SD \cite{cheng2024gin}, assigning attention to nodes for source detection in incomplete networks, and Kang \textit{et al.} introduced HSGBO \cite{kang2026multi}, which leverages a Bayesian-inspired optimization framework handling propagation uncertainty. 

Despite these advantages, a common limitation is the implicit assumption of a strictly causal relationship between observed snapshots and sources, which overlooks the impact of confounding noise such as fake traffic injection or external events, thereby limiting the detection performance. Although Wang \textit{et al.}  attempted to integrate denoising mechanisms into DRSDBFL \cite{wang2024distributed}, the denoising module focuses on node-level noise rather than snapshot-level noise. Consequently, the causal entanglement problem within observations remains a significant challenge for accurate source detection.

\subsection{Source Detection in Hypergraphs}

As an extension of pairwise graphs, hypergraphs exhibit powerful graph modeling capabilities for capturing higher-order relationships and have demonstrated significant efficacy across various domains \cite{DBLP:conf/nips/ZhouHS06, DBLP:conf/mm/Wang0024}. To leverage these topological advantages, recent works \cite{feng2019hypergraph, DBLP:journals/pr/BaiZT21} have developed diverse message aggregation strategies for high-order representations. HGNN \cite{feng2019hypergraph} pioneered this direction by modeling hypergraphs via clique expansion prior to embeddings, whereas UniGNN \cite{DBLP:conf/ijcai/HuangY21} employed star expansion for hypergraph learning. Accordingly, Chien \textit{et al.} \cite{DBLP:conf/iclr/ChienPPM22} introduced a generalized framework that implements hypergraph message passing as multiset functions. Despite these advancements, efficiently modeling diffusion dynamics within hypergraphs remains a critical challenge. Although Xia \textit{et al.} \cite{xia2022self} have attempted to integrate hypergraph structures into the Transformer architecture, this approach inevitably introduces computational overhead, thereby limiting its practical efficiency.

By explicitly capturing high-order interactions \cite{antelmi2023survey}, hypergraph snapshots extract more expressive propagation features than those of pairwise graphs \cite{cai2022hypergraph}. Currently, few works have explored source detection approaches within hypergraphs. Among these studies, Cheng \textit{et al.} proposed SourceDetMamba, employing a graph-aware state space model to infer sources from snapshot sequences \cite{cheng2025sourcedetmamba}, while He \textit{et al.} treated disease source tracing as a heterogeneous hypergraph problem, using a heterogeneous hypergraph attention network to capture diverse modes of propagation \cite{he2025hhan}. Nevertheless, the majority of the existing source detection methods still depend on pairwise networks. By explicitly characterizing the group interactions in social networks, hypergraphs generate more complex diffusion dynamics than pairwise networks, rendering the inverse inference process prone to causal entanglement. Therefore, accurate source detection in hypergraphs remains a critical challenge.

\subsection{Causal Disentanglement}
Causal disentanglement aims to decouple mutually independent latent variables from observational data, with each reflecting a specific underlying causal mechanism. Recently, dual-branch architectures have emerged as an effective approach for causal disentanglement, explicitly partitioning the representation into two isolated spaces corresponding to causal factors and non-causal noise. Liu \textit{et al.} \cite{DBLP:conf/cvpr/LiuCSBA22} introduced a dual-encoder framework to separately model causal and non-causal features for motion forecasting, while Wang \textit{et al.} \cite{DBLP:conf/mm/0008CHYWY023} utilized dual-branch causal mechanisms to explicitly decouple task-specific semantic features from domain bias in unsupervised domain adaptation. Given that dual-branch structures provide an effective mechanism for isolating distinct latent factors, they are suitable to address the signal-noise entanglement problem in propagation snapshots. Inspired by previous studies, we develop a refined dual-branch variational encoder to explicitly separate causal propagation semantics for source localization.



\section{Problem Definition}
\paragraph{Hypergraph Structure.}
Let $\mathcal G = (\mathcal V, \mathcal{E})$ represent the hypergraph, where $\mathcal V$ and $\mathcal{E}$ denote the node set and hyperedge set, respectively. Each hyperedge $e \in \mathcal{E}$ contains a subset of nodes from $\mathcal V$. The structural relationships of the hypergraph are captured by the incidence matrix $H \in \{0, 1\}^{N \times M}$, which is defined as:
\begin{equation}
H_{ij} =
\begin{cases}
1, & \text{if } v_i \in e_j \\
0, & \text{otherwise}.
\end{cases}
\end{equation}

Given the degree of node $v_i$ as $d_{v}(i) = \sum_{j=1}^{M} H_{ij}$ and the degree of hyperedge $e_j$ as $d_{e}(j) = \sum_{i=1}^{N} H_{ij}$, the normalized hypergraph Laplacian is  $L = I - D_v^{-1/2} HD_e^{-1} H^\top D_v^{-1/2}$, where $D_v$ and $D_e$ denote the degree matrices of nodes and hyperedges, respectively \cite{benson2016higher}.

\paragraph{Snapshot-based Source Detection.}
Given a sequence of observed snapshots $O = \{ O_1, \ldots, O_K \}$ on a network $\mathcal G$, classical snapshot-based source localization aims to infer the unknown source set $S \subseteq \mathcal V$ that initiates the diffusion process under a predefined propagation model. This problem is typically formulated as an inverse inference task by estimating the most likely source set:
\begin{equation}
    \hat{S}=\arg \max _{S \subseteq V} P(O \mid S, \mathcal G)
\end{equation}
However, diffusion processes are inevitably perturbed by external noise, which entangles the causal propagation signals within the observed snapshots, thereby incurring performance degradation.


\paragraph{HyperCD Process for Source Detection.}
To address the causal entanglement problem, we propose the HyperCD framework for accurate source localization in hypergraphs, which tackles the task through a three-step procedure, comprising causal disentanglement, temporal reversal, and hypergraph-aware inference.

\textit{1) Causal Disentanglement.} To mitigate the causal entanglement problem, we introduce a disentanglement function $\mathcal{D}_{\Theta}(\cdot)$ that maps the observed snapshot sequence $O$ into two independent latent spaces corresponding to causal propagation semantics and confounding environmental noise, which is formulated as:
\begin{equation}
    (Z_{p}, Z_{e}) = \mathcal{D}_\Theta(\mathbf{O})
\end{equation}
where $\Theta$ represents the learnable parameters of the dual-branch encoder, $Z_{p}$ and $Z_{e}$ denote the causal propagation branch and the confounding noise branch, respectively.

\textit{2) Temporal Reversal.} To align the temporal sequence with the inverse inference process, we apply a reverse operator $\tau(\cdot)$ to the disentangled propagation embeddings.
\begin{equation}
    \tilde{Z}_{p} = \tau(Z_{p})
\end{equation}

\textit{3) Hypergraph-aware Inference.} To model high-order diffusion dynamics in hypergraphs, we introduce an attention-based hypergraph aggregation strategy followed by a scoring function to estimate source likelihoods, formulated as:
\begin{equation}
    \hat S = g_\theta(\mathcal{A_H}(\tilde{Z}_{p}))
\end{equation}
where $\hat S$ represent the estimated sources, $\mathcal{A_H}(\cdot)$ denotes attention-based hypergraph aggregation mechanisms, and $g(\cdot)$ is the scoring function with learnable parameters $\theta$.

\paragraph{Discussion.} Accurately inferring the sources $S^*$ from the discrete observation sequence $O$ presents critical technical challenges. A primary issue is causal entanglement. Since the diffusion states in $O$ are jointly shaped by the true source propagation and unpredictable environmental noise, directly mapping these entangled features to the source nodes often leads to biased inferences. Moreover, the high-order dependencies in hypergraphs introduce complex diffusion dynamics, making the inverse diffusion reconstruction highly challenging. Consequently, it is essential to decouple the causal propagation semantics and capture the high-order diffusion dynamics for source detection.

\section{Method}
\begin{figure*}[t]
\centering
\includegraphics[width=\textwidth]{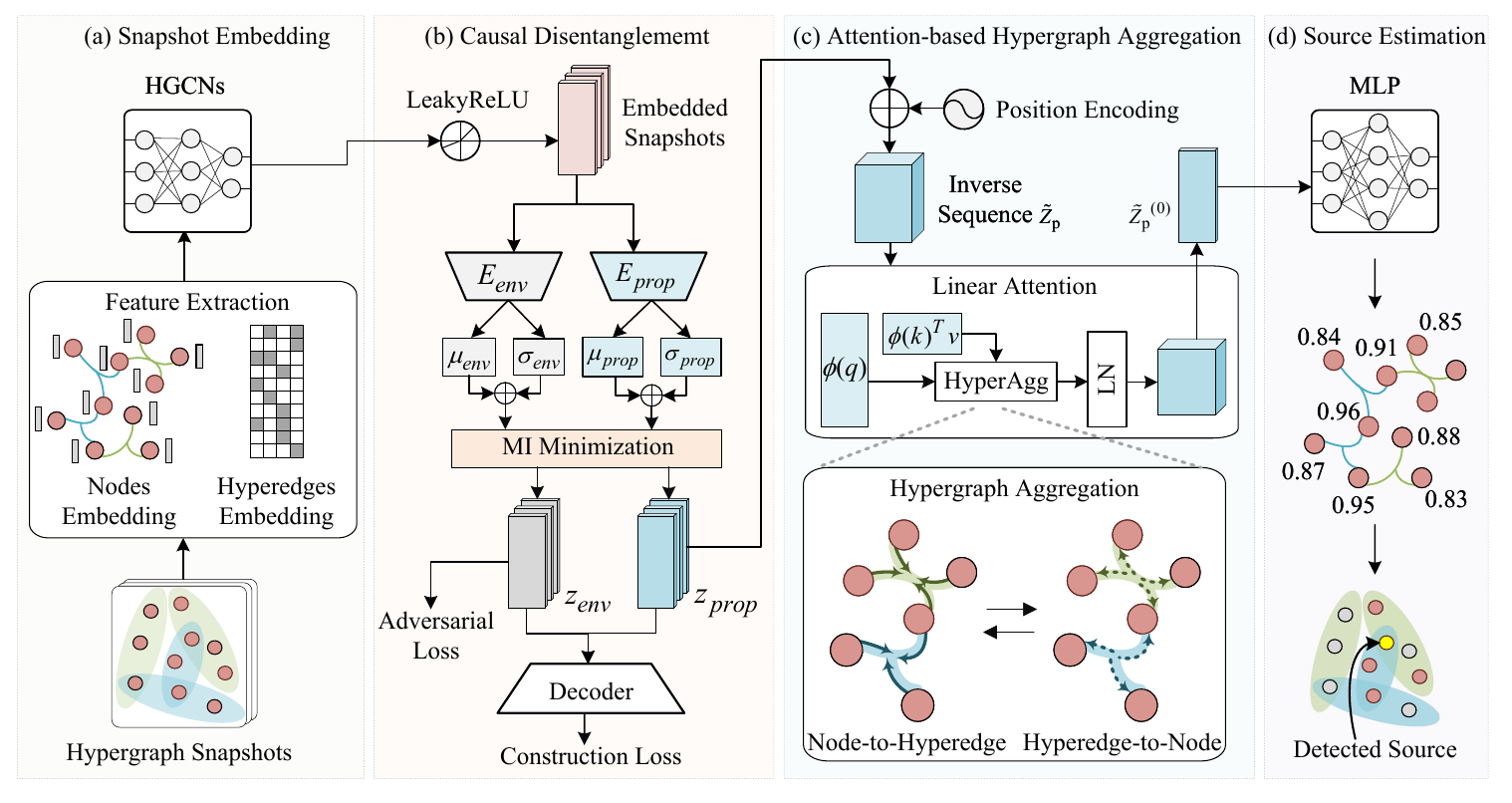}
\caption{Schematic of the proposed HyperCD framework. (a) Feature Construction: For initial feature embeddings, individual infection states and structural positions are aggregated to construct node features, which are subsequently processed by HGCNs. (b) Causal Disentanglement: A dual-branch variational autoencoder is devised, with a disentanglement loss term constraining the module to disentangle causal propagation semantics from confounding noise. (c) Attention-based Hypergraph Aggregation Module: Based on the decoupled propagation branches, an attention-based hypergraph aggregation module is developed, which alternates between node-to-hyperedge and hyperedge-to-node propagation to capture complex diffusion dynamics of hypergraphs. (d) Output: Learned representations are fed into an MLP to infer the sources.}
\label{overview}
\end{figure*}

In this section, we present the HyperCD framework, as illustrated in Figure~\ref{overview}, which decouples propagation semantics and captures high-order diffusion for accurate source detection in hypergraphs. The framework consists of a feature construction module, followed by two key components, namely the causal disentanglement module and the attention-based hypergraph aggregation module.

\subsection{Feature Construction}
To characterize the discrete snapshots, we construct the initial node features by integrating propagation dynamics and structural information. For node $v$, its propagation dynamics are encoded by a binary infection state $\mathbf X_v^1 \in \{+1,-1\}$ and a infection timestamp $\mathbf X_v^2 \in \{t_{v_i},-1\}$. To further capture the topological significance of nodes within the infected subgraph $\mathcal{G}_+$, a Laplacian positional encoding $\mathbf{X}_v^3$ is incorporated into feature construction. Specifically, through the eigendecomposition $L=U \Lambda U^T$, we extract $k$ non-trivial eigenvectors to construct $ \mathbf{X}_v^3 = U_{:,2:k+1}$. Consequently, the unified node representation is obtained by concatenating these components, formulated as:
\begin{equation}
    \mathbf X_v=\left[\|_{i=1}^{d} \mathbf X^i_v\right]
\end{equation}
where $\|$ denotes the concatenation operator, $\mathbf X_v$ represent the unified embedding of node $v$, and $d=3$ represents the dimension.

To capture the high-order dependencies, we integrate the structure of the hypergraph into the initial embeddings. Based on the node embeddings $\mathbf X\in \mathbb{R}^{N \times d}$ and the incidence matrix $H$, we employ the Hypergraph Convolutional Networks (HGCNs) to generate the corresponding structure-aware representations of snapshots, which are updated as:
\begin{equation}
    \mathbf{X^{(l+1)}}= \text{ReLU}\left(\mathbf{D}_v^{-\frac{1}{2}} H \mathbf{D}_e^{-1}H^{\top} \mathbf{D}_v^{-\frac{1}{2}} \mathbf{X^{(l)}}\mathbf{W}\right)
\end{equation}
where $\text{ReLU}(\cdot)$ is the LeakyReLU activation function and $\mathbf W$ represents the learnable parameter matrix.


\subsection{Causal Disentanglement}
To decouple causal propagation semantics, we design a causal disentanglement module, comprising a causal–noise disentangling encoder and a disentanglement-guided generator.

\subsubsection{Causal–Noise Disentangling Encoder.} To explicitly decouple the underlying causal propagation semantics from confounding noise, we design the causal–noise disentangling encoder via a dual-branch architecture. Specifically, the encoder branches $E_{prop}$ and $E_{env}$ map the feature embeddings into the propagation latent space and the noise latent space, respectively. In each branch, the encoders learn the respective mean vectors $\mu$ and variance vectors $\sigma$, via two independent linear projections:
\begin{equation}
    \begin{split} 
    \mu_p = X W_\mu^1 + b_\mu^1, \quad \log \sigma_p^2 = X W_\sigma^1 + b_\sigma^1
    \\
    \mu_e = X W_\mu^2 + b_\mu^2, \quad \log \sigma_e^2 = X W_\sigma^2 + b_\sigma^2
    \end{split}
\end{equation}
where $W$ and $b$ denote the learnable weight matrix and bias. $\mu_p$ and ${\sigma}_{p}$ characterize the learned parameters for the propagation latent space, while ${\mu}_{e}$ and ${\sigma}_{e}$ characterize those for the confounding noise latent space. Subsequently, we sample the propagation branch $z_{p}$ and the confounding noise branch $z_{e}$ by applying the reparameterization trick, formulated as:
\begin{equation}
\begin{aligned}
    \mathbf{z}_{p} = {\mu}_{p} + \exp(\frac{1}{2} \log {\sigma}_{p}^2) \odot {\epsilon}\\
    \mathbf{z}_{e} = {\mu}_{e} + \exp(\frac{1}{2} \log {\sigma}_{e}^2) \odot {\epsilon}
\end{aligned}
\end{equation}
where ${\epsilon} \sim \mathcal{N}(0, {I})$ denotes a standard Gaussian noise for reparameterization. The representation of decoupled branches is:
\begin{equation}
    \mathbf{Z}_p = \{\mathbf{z}_p{(t)}\}_{t=1}^T \quad \mathbf{Z}_e = \{\mathbf{z}_e{(t)}\}_{t=1}^T
\end{equation}
where $\mathbf{Z}_p$ and $\mathbf{Z}_e$ denote the propagation branch and noise branch, respectively, and $T$ is the length of the snapshot sequence.
\subsubsection{Disentanglement-Guided Generator.} 
To verify the reconstruction quality of the branches, we employ the disentanglement-guided generator $D_{\psi}$ to reconstruct the structural embeddings in the input space. Specifically, giving the joint representation $\mathbf{Z} = [\mathbf{Z}_p \parallel \mathbf{Z}_e]$, the decoder employs a two-layer mapping, formulated as:
\begin{equation}
    \hat{\mathbf{X}} = \text{ReLU} \left(\mathbf{Z}\mathbf{W}_1 + \mathbf{b}_1 \right)\mathbf{W}_2 + \mathbf{b}_2
\end{equation}
where $W_1$ and $W_2$ are the learnable weight matrices, and $b_1$ and $b_2$ are corresponding bias vectors. By comparing the reconstructed output $\hat{\mathbf{X}}$ with $\mathbf{X}$, the reconstruction constraint guides both branches to retain representative information from the snapshot embedding.

\subsection{Attention-based Hypergraph Aggregation} 
In this section, we design the attention-based hypergraph aggregation module to capture diffusion dynamics in hypergraphs, consisting of reverse sequence encoding, efficient linear attention, hypergraph message aggregation, and source likelihood estimation.

\subsubsection{Reverse Sequence Encoding.} To capture the position-aware reverse sequence for source detection, we apply a positional encoding layer that incorporates positional information into the disentangled snapshot embeddings. For the \textit{t}-th snapshot representation, the positional encoding layer is defined as:
\begin{equation}
    \mathbf{\tilde z}_p{(t)} = \mathbf{z}_p{(t)} + {\tau}{(t)}
\end{equation}
where ${\tau}{(t)}$ is the associated timestamp of embedding $\mathbf{z}_p{(t)}$. Subsequently, the position-aware embeddings are arranged in reverse order to guide the inverse inference process from the terminal observations to the initial source, formulated as:
\begin{equation}
\label{1}
    \mathbf{\tilde Z}_p=[\mathbf{\tilde z}_p{(T)}, \dots, \mathbf{\tilde z}_p{(1)}]
\end{equation}

\subsubsection{Efficient Linear Attention.} Transformer-based attention effectively captures long-range dependencies and global context, making it well-suited for inferring sources. However, the computational overhead significantly limits efficiency in large-scale networks \cite{chen2021litegt}. Therefore, we develop a linear hypergraph attention mechanism to alleviate the complexity by avoiding the explicit computation of pairwise attention scores, efficiently capturing the propagation dynamics in the hypergraph. Specifically, the attention computation exploits the associativity of matrix multiplication, enabling an approximation of global attention with linear computational complexity of $\mathcal{O}(N)$ \cite{xia2022self}. Given queries $\mathbf{Q}$, keys $\mathbf{K}$, and values $\mathbf{V}$ derived from input features, the linear attention mechanism is formulated by substituting the softmax operator with a kernel function $\phi(\mathbf x)=\text{elu}(\mathbf x)+1$, which acts as a positive feature map to achieve a low rank factorization of the global relational structure. To ensure the numerical stability, the linear attention operator $\mathcal A(\cdot)$ is formulated as:
\begin{equation} 
\mathcal{A}(\mathbf{q}, \mathcal{S}) = \frac{\phi(\mathbf{q}) \sum_{(\mathbf{k}, \mathbf{v}) \in \mathcal{S}} \left( \phi(\mathbf{k})^\top \mathbf{v} \right) }{\phi(\mathbf{q}) \sum_{(\mathbf{k}, \mathbf{v}) \in \mathcal{S}} \phi(\mathbf{k})^\top} 
\end{equation}
where $\mathbf q$ denotes the query vector for a target node or hyperedge, and $\mathcal S$ represents the set of corresponding key-value pairs. 

\subsubsection{Hypergraph Message Aggregation.} To capture the high-order structural dependencies in hypergraphs, we develop a hypergraph message aggregation strategy, consisting of node-to-hyperedge propagation and hyperedge-to-node propagation processes.

\paragraph{Node-to-Hyperedge Propagation.}
Let $\mathcal{N}(j) = \{\mathbf v_i \in \mathcal{V} \mid (\mathbf{v_i}, \mathbf{e_j}) \in H\}$ represent the set of nodes incident to hyperedge $e_j$, and $\mathcal{I}(i) = \{\mathbf{e_j} \in \mathcal{E} \mid (\mathbf{v_i}, \mathbf{e_j}) \in H\}$ denote the set of hyperedges containing node $v_i$. Specifically, the node-to-hyperedge propagation aggregates incident node features via the incidence matrix to update hyperedge representations, which is formally presented as:
\begin{equation}
\begin{aligned}
\mathbf{\tilde Z}^{n\rightarrow e,l-1} &= \big\|_{j=1}^{M}\mathbf{\tilde z}_j^{n\rightarrow e,l-1} \\
\mathbf{\tilde z}_j^{n\rightarrow e,l-1} &= \mathcal A\left (\mathbf{q_j},\left(\mathbf{k}_i,\mathbf{u}_i\right)_{i \in {\mathcal N}\left( j\right)}\right)
\end{aligned}
\end{equation}
where $\mathbf{\tilde z_j^{n\rightarrow e,l-1}}$ denotes the aggregated hyperedge features for hyperedge $e_j$; $\mathcal A(\cdot)$ indicates the aggregation operator; $\mathbf q_j$, $\mathbf k_i$, and $\mathbf u_i$ denote the query, key and value vectors for hyperedge $e_j$ and node $v_i$, respectively.  The vectors are computed via linear projections as:
\begin{equation}
    \mathbf {q_j} = \mathbf{\tilde E}_j^{l-1}W_Q \quad \mathbf k_i=\mathbf{\tilde Z}_i^{l-1}W_K \quad \mathbf u_i=\mathbf{\tilde Z}_i^{l-1}W_U
\end{equation}
Here, $\mathbf{\tilde E}^{l-1}_j \in R^{E\times d}$ and $\mathbf{\tilde Z}_i^{l-1}$ represent the embedding of hyperedge $e_j$ and node $v_i$ at the $l-1$ layer, respectively. $\mathbf{W}_Q, \mathbf{W}_K$, and $\mathbf{W}_U \in \mathbb{R}^{d \times d}$ are learnable weight matrices.

\paragraph{Hyperedge-to-Node Propagation.}
Subsequently, the hyperedge-to-node propagation updates the node representations by aggregating messages from the incident hyperedges $\mathbf{\tilde Z}^{n\rightarrow e,l-1}$, formulated as:
\begin{equation}
    \begin{aligned}
        \mathbf{\tilde Z}^{e\rightarrow n,l} &= \big\|_{i=1}^{N}\mathbf{\tilde z}_i^{e\rightarrow n,l} \\ 
        \mathbf{\tilde z}_i^{e\rightarrow n,l} &= \mathcal A\left (\mathbf{q'_i},\left(\mathbf{k'_j,u'_j}\right)_{j \in {\mathcal I}\left(i\right)}\right)
    \end{aligned}
\end{equation}
where $\mathbf{\tilde z}_i^{e\rightarrow n,l}$ represents the updated node feature for node $v_i$. Based on a node-hyperedge importance symmetry assumption, the attention computation shares most parameters with the aforementioned propagation. Accordingly, the query and key vectors are swapped, and the vectors are formulated as:
\begin{equation}
    q_i' = k_i \quad k_j'=q_j \quad u_j'=\mathbf{\tilde Z}^{n\rightarrow e,l-1}W_U
\end{equation}


\subsubsection{Source Likelihood Estimation.}
After $L$ attention layers, the learned representation $\mathbf{\tilde Z}$ encodes the reversed temporal dynamics of the cascade. To detect the diffusion source, we extract $\mathbf{\tilde Z}^{(1)} \in \mathbf{\tilde Z}$, which corresponds to the initial diffusion state. This element is subsequently fed into a Multi-Layer Perceptron (MLP) with a sigmoid activation to estimate the node-wise source likelihoods, formulated as:
\begin{equation}
    \hat{\mathbf{Y}} = \text{ReLU}\left(\text{MLP}(\mathbf{\tilde Z}_{p}^{(1)})\right)
\end{equation}
where $\hat{\mathbf{Y}}$ represents the probability of each node being the source.

\subsection{Loss Function}
Source detection is formulated as a binary node classification problem in our proposed framework. To mitigate the class imbalance between sources and non-sources, a cross-entropy loss  is computed utilizing the propagation representations $Z_p$:
\begin{equation}
\mathcal{L}_{\text{inf}} = - \sum_{v \in S} \log(\hat{y}_{v}) - \alpha \sum_{v \in \mathcal{V} \setminus S} \log(1 - \hat{y}_{v})
\end{equation}
where $S$ and $\mathcal{V}$ denote the source set and total nodes set, respectively, $\hat{y}_v$ is the predicted source probability, and $\alpha = \frac{|\mathcal{S}|}{|\mathcal{V}| - |\mathcal{S}|}$ acts as a balancing coefficient. Similarly, an auxiliary inference loss $\mathcal{L}_{aux}$ is derived from the confounding noise representations $Z_e$.

Consequently, to enforce the semantic separation between the latent branches, an adversarial inference constraint is introduced:
\begin{equation}
    \mathcal L_{AD} = \frac{\mathcal L_{inf}}{\mathcal L_{aux}}
\end{equation}
This constraint maximizes source identifiability within the propagation branch while suppressing it in the noise branch. Moreover, a Jensen-Shannon estimator is employed to compute the mutual information loss $\mathcal{L}_{MI}$, which penalizes semantic overlap to guarantee statistical independence between $Z_p$ and $Z_e$. By integrating the standard evidence lower bound (ELBO) loss $\mathcal{L}_{LB}$ to maintain the validity of causal reconstruction, the overall objective function is:
\begin{equation}
    \mathcal{L} = \mathcal{L}_{inf} + \lambda_1\mathcal{L}_{LB} + \lambda_2\mathcal{L}_{MI} + \lambda_3\mathcal{L}_{AD}
\end{equation}
where $\{\lambda_i\}_{i=1}^3$ are tunable hyperparameters that dictate the relative importance of each optimization objective.

\section{Experiments}
\subsection{Experimental Setup}
\subsubsection{Datasets.} To validate the effectiveness of HyperCD, experiments are conducted on six datasets, including Zoo \cite{zoo_111}, House \cite{chodrow2021generative}, NTU2012 \cite{chen2003visual}, Cora \cite{mccallum2000automating}, Citeseer \cite{leskovec2012learning}, and PubMed \cite{yadati2019hypergcn}.

\subsubsection{Baselines.}
The proposed approach is compared against eight competitive baselines. These methods include snapshot-based models such as GCNSI \cite{dong2019multiple}, SL-VAE \cite{ling2022source}, IVGD \cite{wang2022invertible}, GIN-SD \cite{cheng2024gin}, and DRSDBFL \cite{wang2024distributed}. Additionally, the model is evaluated against HHAN \cite{he2025hhan}, HyperDet \cite{cheng2025hyperdet}, and SDMamba \cite{cheng2025sourcedetmamba}, which represent advanced methods for source detection in hypergraphs.

\subsubsection{Metrics.}
Four metrics are employed to evaluate performance: Accuracy (ACC), F-score, Area Under the ROC Curve (AUC), and Precision@20 (P@20). ACC measures the overall proportion of correctly identified samples, F-score provides a balanced assessment by jointly considering precision and recall, and AUC reflects the model's ability to rank true sources higher than non-source nodes across various thresholds. Furthermore, P@20 evaluates the precision among the top 20 ranked nodes.

\subsubsection{Implementation.}
Contagion dynamics over the hypergraph are modeled by randomly selecting a subset of nodes, with a proportion $\rho_s$, as the initial sources. The infection probability $p$ for each node follows a bounded uniform distribution whose range is controlled by the infection rate, and the hyperedge propagation probability scales linearly with the number of infected nodes within the hyperedge. To simulate confounding noise during the diffusion process, the propagation probabilities of both nodes and hyperedges are further perturbed by an additive zero-mean Gaussian term with noise intensity $\eta$. The initial observed snapshot is recorded when the infection ratio reaches 10\%, followed by additional snapshots collected whenever the infection ratio increases by a fixed increment $\delta$.

\subsection{Performance Comparison and Analysis}
\subsubsection{Source Detection Performance}
We evaluate the source detection performance of HyperCD across six datasets under the Susceptible-Infected (SI) \cite{yang2020locating} diffusion model, with the results compared against state-of-the-art baselines as summarized in Table~\ref{sota}. Among pairwise interaction-based source detection baselines, IVGD and SL-VAE outperform GCNSI, highlighting the value of dynamic propagation priors. Regarding approaches that focus on source detection noise, DRSDBFL shows marginal improvements via partial node denoising, while GIN-SD leverages node-wise attention to handle snapshot incompleteness. Among baselines with high-order interactions, HHAN, HyperDet, and SDMamba are the most prominent recent studies. Remarkably, compared to the best-performing baseline, HyperCD achieves an average F-score improvement of 7.79\% across all datasets. This substantial gain is attributed to the proposed causal disentanglement strategy, which decouples essential propagation semantics from noise within the snapshots, thereby enhancing the effectiveness of the framework. Furthermore, the disentanglement loss with mutual information minimization effectively enforces semantic decoupling, while the stacked linear attention adequately captures the complex diffusion dynamics of hypergraphs for accurate source detection.

\begin{table*}[t]
    \centering
    \caption{Performance comparison with state-of-the-art methods across six datasets.}
    \renewcommand{\arraystretch}{1.0}
  \setlength{\tabcolsep}{3pt}
  \resizebox{\textwidth}{!}{
        \begin{tabular}{lcccccccccccccccccc}
        \toprule
        \label{sota}
        \multirow{2}{*}{Method} & 
        \multicolumn{3}{c}{Zoo} & \multicolumn{3}{c}{House} & \multicolumn{3}{c}{Citeseer} & 
        \multicolumn{3}{c}{NTU} & \multicolumn{3}{c}{Cora} & \multicolumn{3}{c}{Pubmed} \\
        
        \cmidrule(lr){2-4} \cmidrule(lr){5-7} \cmidrule(lr){8-10} 
        \cmidrule(lr){11-13} \cmidrule(lr){14-16} \cmidrule(lr){17-19}
        
         & ACC & F-score & AUC & ACC & F-score & AUC & ACC & F-score & AUC & ACC & F-score & AUC & ACC & F-score & AUC & ACC & F-score & AUC \\
        \midrule
        
        GCNSI   & 0.823 & 0.372 & 0.810 & 0.828 & 0.329 & 0.811 & 0.805 & 0.237 & 0.801 & 0.817 & 0.306 & 0.825 & 0.802 & 0.347 & 0.807 & 0.816 & 0.279 & 0.823 \\
        IVGD    & 0.847 & 0.589 & 0.835 & 0.824 & 0.529 & 0.836 & 0.824 & 0.529 & 0.811 & 0.859 & 0.532 & 0.856 & 0.810 & 0.547 & 0.823 & 0.819 & 0.563 & 0.817 \\
        SL-VAE  & 0.851 & 0.602 & 0.830 & 0.842 & 0.531 & 0.824 & 0.824 & 0.557 & 0.828 & 0.846 & 0.529 & 0.839 & 0.824 & 0.525 & 0.830 & 0.824 & 0.557 & 0.824 \\
        DRSDBFL & 0.853 & 0.622 & 0.833 & 0.845 & 0.573 & 0.833 & 0.822 & 0.573 & 0.833 & 0.855 & 0.585 & 0.841 & 0.830 & 0.573 & 0.845 & 0.824 & 0.575 & 0.826 \\
        GIN-SD  & 0.861 & 0.694 & 0.834 & 0.859 & 0.624 & 0.854 & 0.831 & 0.584 & 0.819 & 0.860 & 0.634 & 0.841 & 0.831 & 0.584 & 0.819 & 0.822 & 0.587 & 0.828 \\
        HHAN    & 0.935 & 0.745 & 0.809 & 0.865 & 0.672 & 0.814 & 0.833 & 0.694 & 0.773 & 0.865 & 0.672 & 0.814 & 0.955 & 0.794 & 0.809 & 0.855 & 0.663 & 0.803 \\
        HyperDet& 0.901 & 0.792 & 0.892 & 0.884 & 0.726 & 0.875 & 0.869 & 0.748 & 0.882 & 0.877 & 0.715 & 0.869 & 0.915 & 0.807 & 0.923 & 0.870 & 0.723 & 0.871 \\
        SDMamba & 0.915 & 0.797 & 0.920 & 0.938 & 0.836 & 0.941 & 0.905 & 0.795 & 0.897 & 0.947 & 0.869 & 0.944 & 0.931 & 0.813 & 0.922 & 0.905 & 0.784 & 0.897 \\
        \textbf{HyperCD} & \textbf{0.950} & \textbf{0.883} & \textbf{0.945} & \textbf{0.945} & \textbf{0.919} & \textbf{0.963} & \textbf{0.940} & \textbf{0.863} & \textbf{0.934} & \textbf{0.955} & \textbf{0.932} & \textbf{0.963} & \textbf{0.945} & \textbf{0.862} & \textbf{0.943} & \textbf{0.933} & \textbf{0.817} & \textbf{0.927} \\
        \bottomrule
        \end{tabular}
    }
\end{table*}

\subsubsection{Performance under Varying Snapshot Intervals}
\begin{figure}[htbp]
\centering
\includegraphics[width=0.48\textwidth]{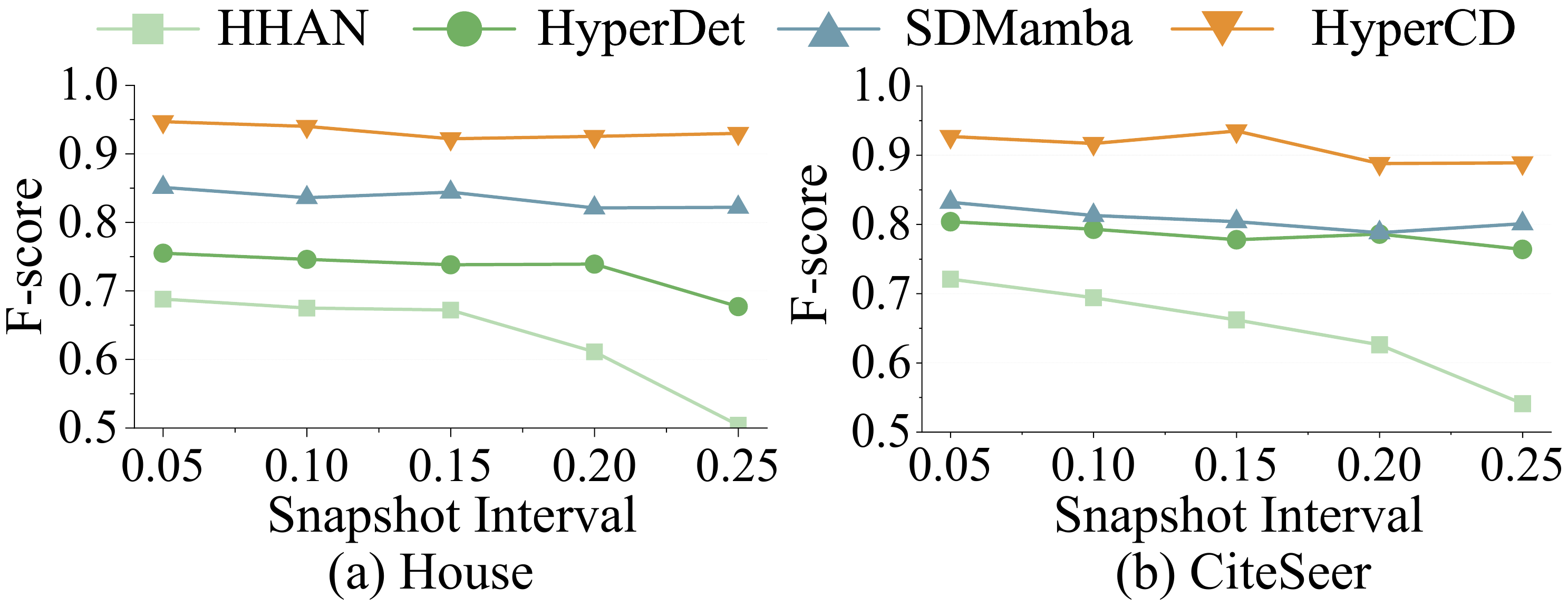}
\caption{Performance comparison across baselines under varying snapshot intervals.}
\label{intv}
\end{figure}

The snapshot interval $\delta$ determines the sparsity of the input snapshot sequence. Specifically, the sampling process starts at a 10\% infection rate, and subsequent snapshots are collected for every $\delta$ increase in the infection proportion until propagation terminates. As in Figure \ref{intv}, the proposed model consistently outperforms the baselines across varying snapshot intervals. This superiority is particularly pronounced at larger intervals, indicating that the model maintains robust performance even when relying on sparse observations.
\begin{figure}[htbp]
\centering
\includegraphics[width=0.48\textwidth]{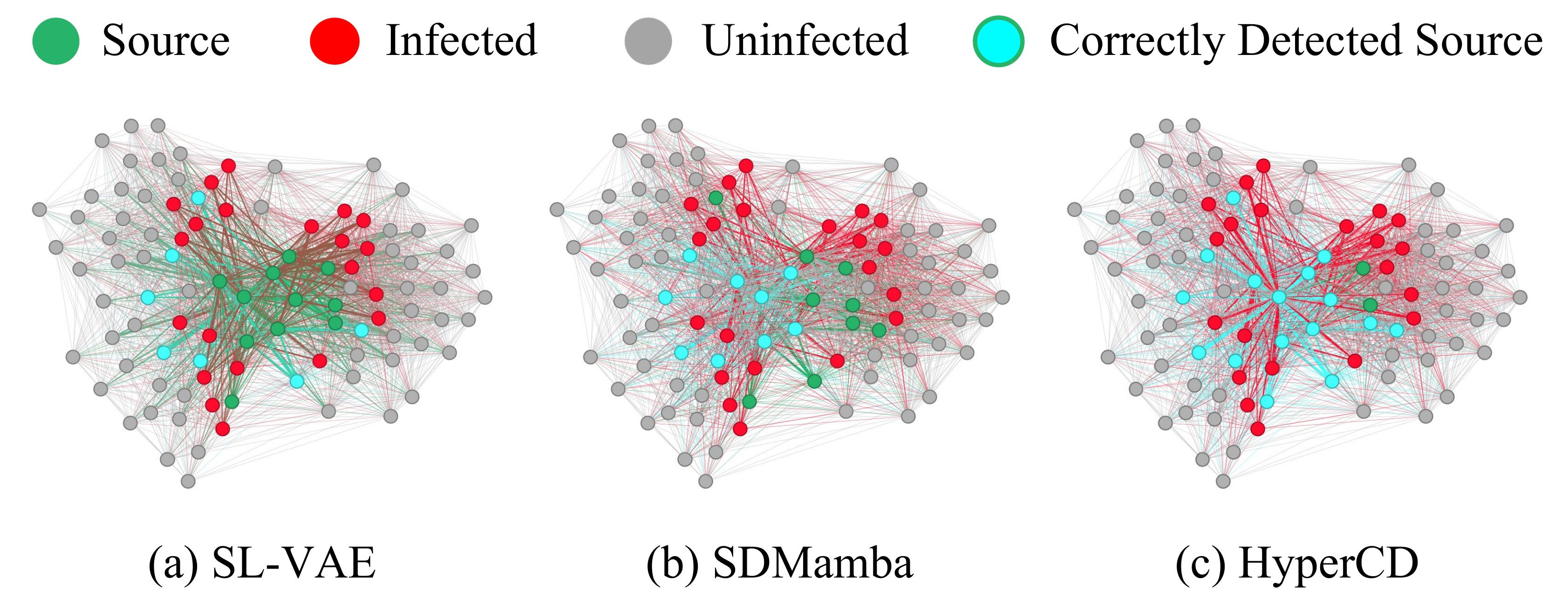}
\caption{Visualization of source detection results on Zoo.}
\label{visu}
\end{figure}

\subsubsection{Visualization}
To provide an intuitive understanding of the source localization performance, we visualize the outputs of HyperCD alongside those of several representative baseline methods on the Zoo dataset. As in Figure \ref{visu}, HyperCD correctly identifies the majority of true source nodes, whereas the representative baselines exhibit relatively noticeable deviations from the ground truth. This observation indicates that HyperCD effectively alleviates the causal entanglement problem and accurately captures hypergraph diffusion dynamics, thereby leading to improved source identification performance compared to other baselines.

\subsection{Ablation Study and Analyses}
\subsubsection{Ablations on Causal Disentanglement Module}
To evaluate the contribution of the Causal Disentanglement Module (CDM) to source detection, we conduct ablation studies on the core components comprising HGCN embedding layers, disentangled encoder layers, and loss terms. Specifically, w/o HGCNs and w/o CDM are established to eliminate the HGCN embedding layers and the entire CDM module, respectively. To assess the necessity of the loss terms, w/o MIloss removes the mutual information loss $\mathcal{L}_{MI}$, while w/o ADloss replaces the adversarial inference loss $\mathcal{L}_{AD}$ with the standard inference loss $\mathcal{L}_{inf}$. Table \ref{abl} demonstrates that removing HGCN embeddings or disentangled encoders incurs performance degradation; notably, the sharp decline observed in w/o CDM highlights the critical importance of the disentanglement mechanism. Moreover, the performance of the loss ablation variants is comparable to that of w/o CDM, suggesting that these loss terms serve as a pivotal constraint for causal disentanglement.

\begin{table}[hbpt]
    \centering
    \caption{Ablation study on causal disentanglement module.}
    \renewcommand{\arraystretch}{0.88}
  \setlength{\tabcolsep}{4.7pt}

  \resizebox{\linewidth}{!}{
        \begin{tabular}{l|cccccc}
        \toprule
        & \multicolumn{2}{c}{House} & \multicolumn{2}{c}{NTU} & \multicolumn{2}{c}{Zoo} \\
        \cmidrule(lr){2-3} \cmidrule(lr){4-5} \cmidrule(lr){6-7}
        Variants  & F-score & P@20 & F-score & P@20 & F-score & P@20\\
        \midrule
        w/o CDM   & 0.842  & 0.764&         0.833 & 0.711 &        0.767 & 0.735\\
        w/o HGCNs   & 0.889  & 0.809&         0.868 & 0.802 &        0.864 & 0.781\\
        w/o ADloss   & 0.850  & 0.775&         0.840 & 0.787 &        0.818 & 0.746\\
        w/o MIloss   & 0.861  & 0.770&         0.863 & 0.781 &        0.825 & 0.764\\
\textbf{HyperCD}&\textbf{0.919} & \textbf{0.883}& \textbf{0.932} & \textbf{0.945}&\textbf{0.883}& \textbf{0.792}\\
        \bottomrule
    \end{tabular}
    }
    \label{abl}
\end{table}
\begin{table}[hbpt]
    \centering
    \caption{Ablation study on the attention-based hypergraph aggregation module.}
    \renewcommand{\arraystretch}{0.88}
    \normalsize
  \setlength{\tabcolsep}{3pt}

  \resizebox{\linewidth}{!}{
        \begin{tabular}{l|cccccc}
        \toprule
        & \multicolumn{2}{c}{House} & \multicolumn{2}{c}{NTU} & \multicolumn{2}{c}{Zoo} \\
        \cmidrule{2-3} \cmidrule{4-5} \cmidrule{6-7}
        Variants  & F-score & time (s) & F-score & time (s) & F-score & time (s)\\
        \midrule
        w/o AHA  & 0.761& 4580.21 & 0.721 & 6228.47 & 0.689 & 818.57 \\
        w/o H-Agg
        & 0.766 & 4667.95 & 0.735 &        6474.12 & 0.717 & 841.00 \\
        w/o L-Att   & 0.911& 6056.58 & 0.928 & 10702.58 & 0.886 & 1891.53\\
        \textbf{HyperCD}&\textbf{0.919} & 4705.92 &\textbf{0.932} & 6523.80 & \textbf{0.883} & 876.40\\
        \bottomrule
    \end{tabular}
    }
    \label{abl2}
\end{table}
\subsubsection{Ablations on Attention-based Hypergraph Aggregation Module}
This section investigates the contribution of the Attention-based Hypergraph Aggregation module (AHA) to source detection. To evaluate the key components of the AHA module, we construct three variants, including w/o AHA, w/o H-Agg, and w/o L-Att. The w/o H-Agg variant removes the hypergraph message aggregation between nodes and hyperedges. The w/o L-Att variant replaces linear attention with standard dot-product attention to assess computational efficiency. As shown in Table \ref{abl2}, performance declines across all removal variants, confirming the necessity of both components within the attention-based hypergraph aggregation module. Notably, although w/o L-Att exhibits a comparable performance, it incurs significantly higher time complexity, highlighting the importance of linear attention for model efficiency.

\begin{figure}[htbp]
\centering
\includegraphics[width=0.48\textwidth]{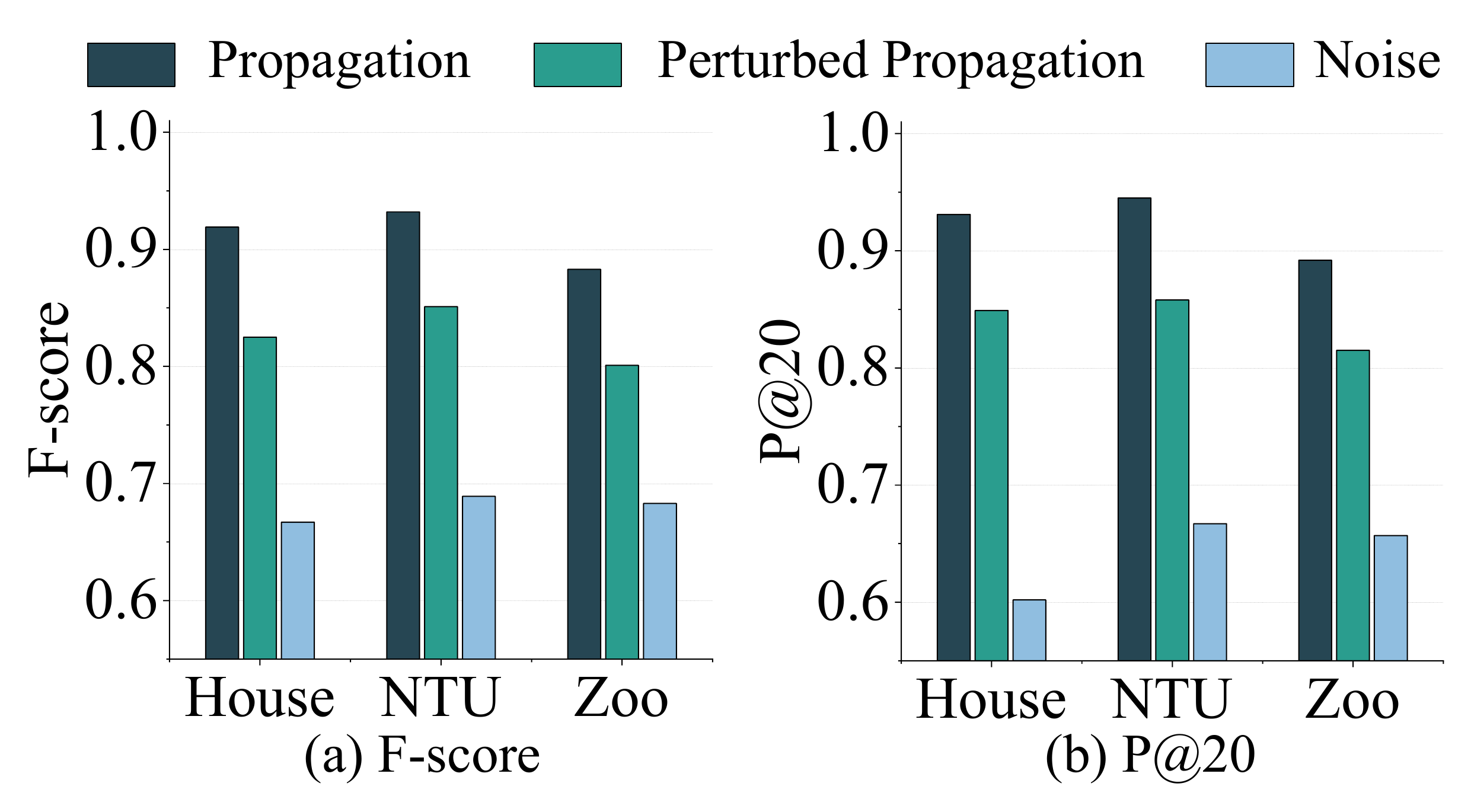}
\caption{Validation of disentanglement effectiveness. Propagation, Perturbed Propagation, and Noise indicate the causal propagation branch, the Gaussian perturbed propagation branch, and the confounding noise branch, respectively. }
\label{disen}
\end{figure}

\subsubsection{Identifiability of Causal Propagation Semantics}
Here, we focus on the decoupled representations $Z_p$ and $Z_e$ to verify the capability of capturing the propagation signals and confounding noise, respectively. To validate this semantic separation, $Z_p$ and $Z_e$ are independently fed into the source inference module. Furthermore, a Gaussian perturbed variant $\tilde{Z}_p$ is introduced to assess the source identifiability of the decoupled propagation branch. As in Figure \ref{disen}, we find that $Z_p$ achieves optimal accuracy, whereas the perturbed $\tilde{Z}_p$ shows significant performance degradation, confirming that $Z_p$ encodes the determinant causal semantics. Moreover, $Z_e$ yields performance comparable to random guessing, verifying that it primarily encodes information irrelevant to source detection and thereby validating the efficacy of the semantic disentanglement.

\subsubsection{Independence of Latent Representations}
\begin{figure}[htbp]
\centering
\includegraphics[width=0.48\textwidth]{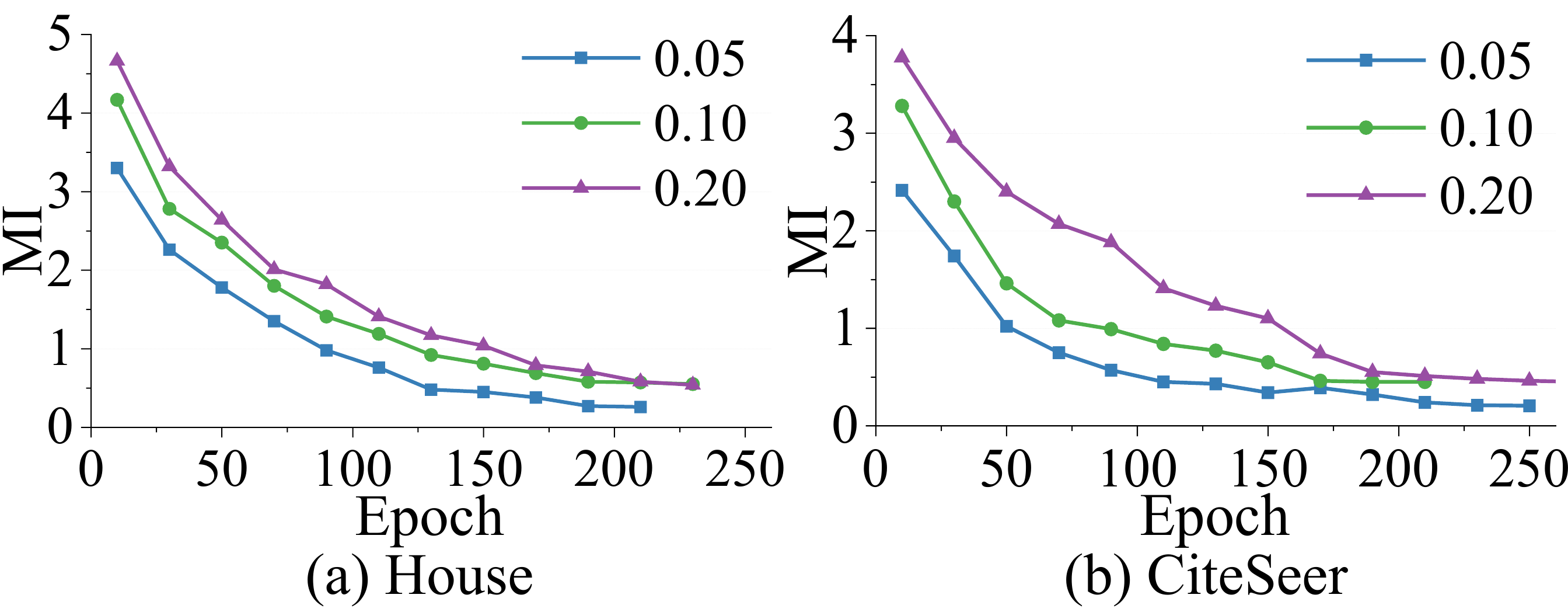}
\caption{Mutual information $MI(Z_p; Z_e)$ over training epochs on House and NTU2012 under different noise intensities.}
\label{mi}
\end{figure}

To quantitatively evaluate the statistical independence of the learned latent representations, we conduct experiments on the House and NTU2012 datasets under Gaussian noise intensities $\eta \in \{0.05, 0.10, 0.20\}$ in the diffusion process. Specifically, we monitor the mutual information $MI(Z_p; Z_e)$ throughout the training process, using the Jensen-Shannon (JS) estimator to quantify the dependence between the propagation semantics $Z_p$ and the confounding noise $Z_e$. As shown in Figure \ref{mi}, the mutual information decreases rapidly during the early epochs and subsequently converges to a near-zero value. This trend indicates that the proposed mutual information minimization objective effectively reduces redundancy, yielding well-disentangled and semantically independent latent representations that play a crucial role in improving source inference performance.

\subsubsection{Impact of Source Ratio}
\begin{figure}[htbp]
\centering
\includegraphics[width=0.48\textwidth]{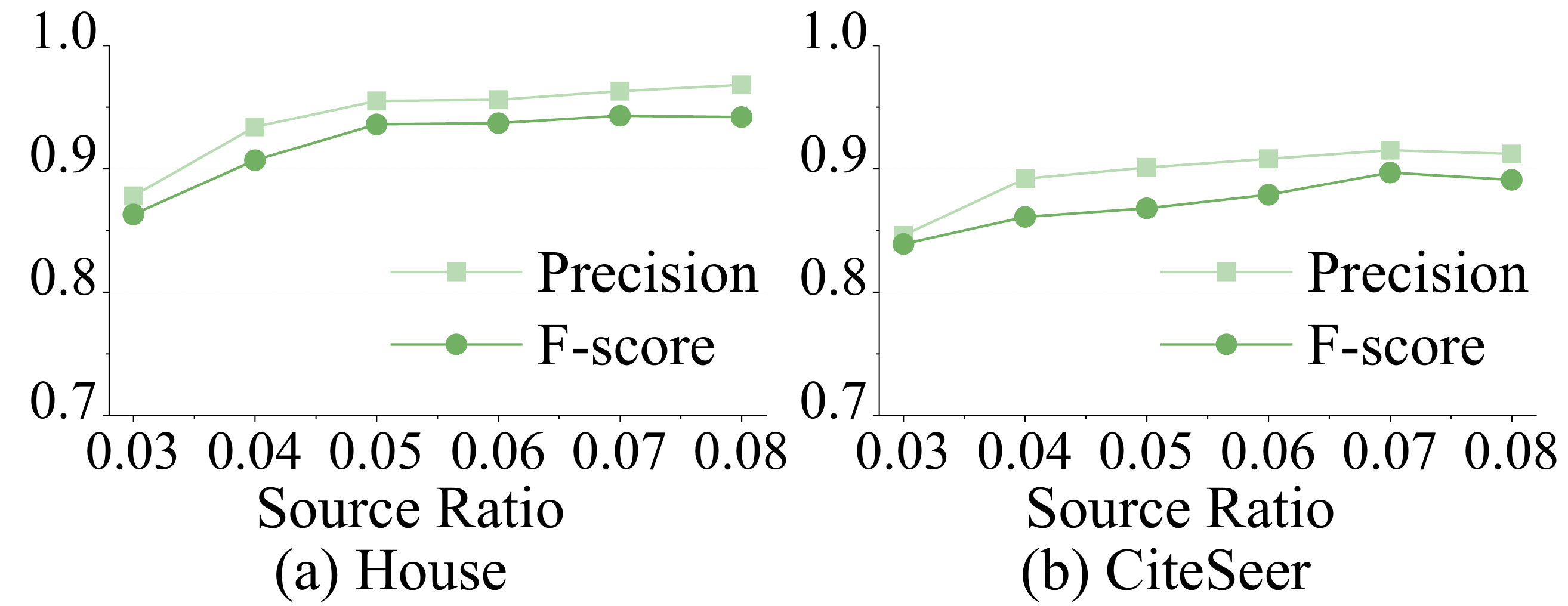}
\caption{Impact of varying the source ratio.}
\label{source}
\end{figure}

The source ratio determines the initial infection density and directly affects the structural observability. Hence, experiments are conducted on House and CiteSeer, with the corresponding results being depicted in Figure~\ref{source}. The source ratio is varied from 0.03 to 0.08, and the performance is evaluated using Precision and F-score metrics. The results reveal that, although slight fluctuations occur at lower source ratios due to the subgraph sparsity, the model remains generally stable. This demonstrates its robustness in handling varying levels of source density.

\subsubsection{Impact of Infection Rate}
\begin{figure}[htbp]
\centering
\includegraphics[width=0.48\textwidth]{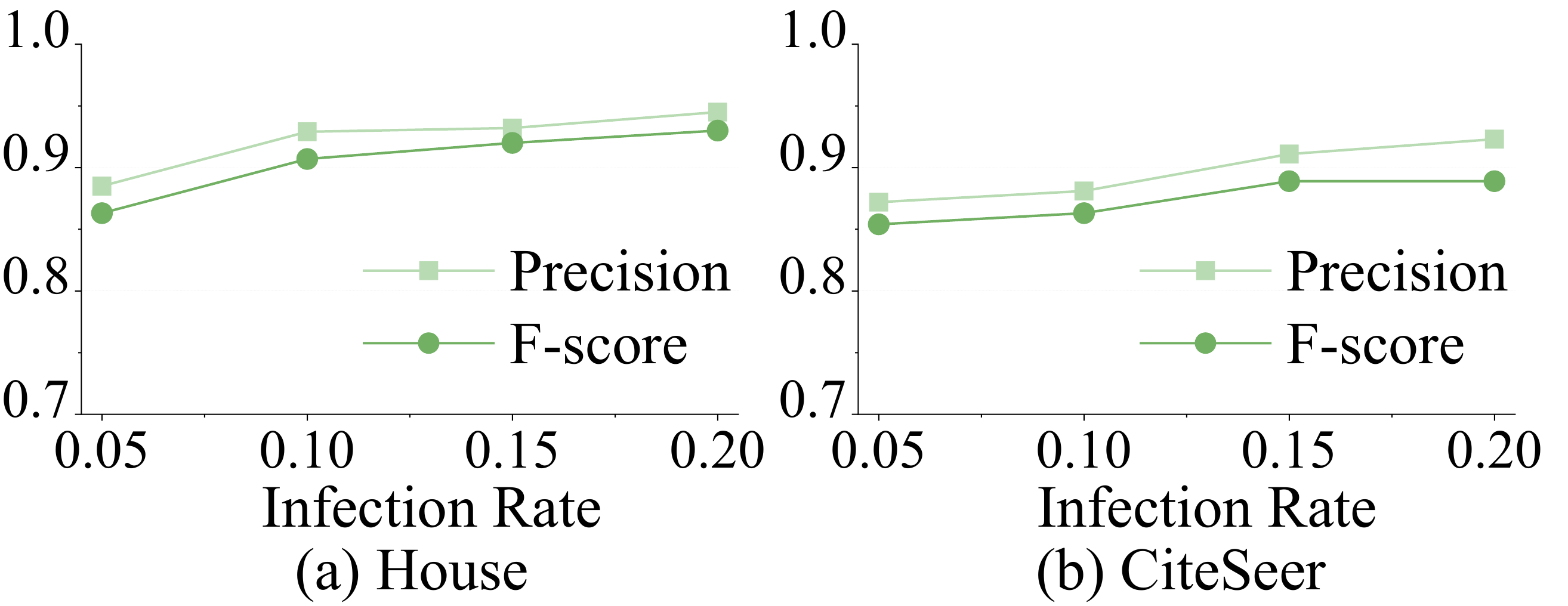}
\caption{Impact of varying the infection rate.}
\label{prop}
\end{figure}

The infection rate governs the propagation speed and thereby influences the temporal diffusion patterns. Consequently, model performance is further analyzed under scenarios with varying infection rates. Specifically, experiments are conducted on the House and Citeseer datasets, with the infection rate ranging from 0.05 to 0.2 in increments of 0.05. The results are depicted in Figure~\ref{prop}, illustrating that our HyperCD maintains consistent effectiveness under varying propagation speeds, demonstrating its adaptability to diverse diffusion dynamics.

\section{Conclusion}
In this paper, we focus on the causal entanglement problem within snapshots and the complex diffusion dynamics in hypergraphs. A source detection approach in \textbf{Hyper}graphs via \textbf{C}ausal \textbf{D}isentanglement (HyperCD) is proposed to effectively and accurately identify diffusion sources from the snapshot sequences. By leveraging disentangled encoders alongside stacked linear hypergraph attention, the model is capable of decoupling causal propagation semantics from confounding noise while accurately modeling hypergraph diffusion dynamics, thereby significantly enhancing overall effectiveness. Extensive experiments across six datasets demonstrate the superiority of our HyperCD compared to baselines. Future work will explore more complex network structures and advanced disentanglement strategies for source detection within high-order networks.

\begin{acks}
This work was supported by the Zhejiang Provincial Excellent Funding for Postdoctoral Researchers under Grant ZJ2025156 and the China Postdoctoral Science Foundation under Grant 2025M784428.
\end{acks}

\bibliographystyle{ACM-Reference-Format}
\bibliography{references}

@article{benson2016higher,
  title={Higher-order organization of complex networks},
  author={Benson, Austin R and Gleich, David F and Leskovec, Jure},
  journal={Science},
  volume={353},
  number={6295},
  pages={163--166},
  year={2016},
  publisher={American Association for the Advancement of Science}
}

@article{majhi2022dynamics,
  title={Dynamics on higher-order networks: A review},
  author={Majhi, Soumen and Perc, Matja{\v{z}} and Ghosh, Dibakar},
  journal={Journal of the Royal Society Interface},
  volume={19},
  number={188},
  pages={43-58},
  year={2022},
  publisher={The Royal Society}
}

@inproceedings{wang2017multiple,
  title={Multiple source detection without knowing the underlying propagation model},
  author={Wang, Zheng and Wang, Chaokun and Pei, Jisheng and Ye, Xiaojun},
  booktitle={Proceedings of the AAAI Conference on Artificial Intelligence},
  pages={279--308},
  year={2017}
}

@inproceedings{dong2019multiple,
  title={Multiple rumor source detection with graph convolutional networks},
  author={Dong, Ming and Zheng, Bolong and Quoc Viet Hung, Nguyen and Su, Han and Li, Guohui},
  booktitle={Proceedings of the 28th ACM International Conference on Information and Knowledge Management},
  pages={569--578},
  year={2019}
}

@inproceedings{ling2022source,
	title={Source localization of graph diffusion via variational autoencoders for graph inverse problems},
	author={Ling, Chen and Jiang, Junji and Wang, Junxiang and Liang, Zhao},
	booktitle={Proceedings of the 30th ACM SIGKDD Conference on Knowledge Discovery and Data Mining},
	pages={1010--1020},
	year={2022}
}

@inproceedings{wang2022invertible,
	title={An invertible graph diffusion neural network for source localization},
	author={Wang, Junxiang and Jiang, Junji and Zhao, Liang},
	booktitle={Proceedings of the ACM Web Conference 2022},
	pages={1058--1069},
	year={2022}
}

@article{cheng2024heuristic,
	title={A heuristic framework for sources detection in social networks via graph convolutional networks},
	author={Cheng, Le and Zhu, Peican and Gao, Chao and Wang, Zhen and Li, Xuelong},
	journal={IEEE Transactions on Systems, Man, and Cybernetics: Systems},
	year={2024},
    volume={54},
    number={11},
    pages={7002-7014},
}

@inproceedings{cheng2024gin,
  title={GIN-SD: source detection in graphs with incomplete nodes via positional encoding and attentive fusion},
  author={Cheng, Le and Zhu, Peican and Tang, Keke and Gao, Chao and Wang, Zhen},
  booktitle={Proceedings of the AAAI Conference on Artificial Intelligence},
  pages={55--63},
  year={2024}
}

@article{wang2024distributed,
  title={Distributed rumor source detection via boosted federated learning},
  author={Wang, Ranran and Zhang, Yin and Wan, Wenchao and Chen, Min and Guizani, Mohsen},
  journal={IEEE Transactions on Knowledge and Data Engineering},
  volume={36},
  number={11},
  pages={5986--6001},
  year={2024},
  publisher={IEEE}
}

@article{yang2020locating,
  title={Locating the propagation source in complex networks with a direction-induced search based Gaussian estimator},
  author={Yang, Fan and Yang, Shuhong and Peng, Yong and Yao, Yabing and Wang, Zhiwen and Li, Houjun and Liu, Jingxian and Zhang, Ruisheng and Li, Chungui},
  journal={Knowledge-Based Systems},
  volume={195},
  number={105674},
  year={2020},
  publisher={Elsevier}
}

@article{antelmi2023survey,
  title={A survey on hypergraph representation learning},
  author={Antelmi, Alessia and Cordasco, Gennaro and Polato, Mirko and Scarano, Vittorio and Spagnuolo, Carmine and Yang, Dingqi},
  journal={{ACM Computing Surveys}},
  volume={56},
  number={1},
  pages={1--38},
  year={2023},
  publisher={ACM New York, NY}
}

@inproceedings{cai2022hypergraph,
  title={Hypergraph Structure Learning for Hypergraph Neural Networks},
  author={Cai, Derun and Song, Moxian and Sun, Chenxi and Zhang, Baofeng and Hong, Shenda and Li, Hongyan},
  booktitle={Proceedings of the 31st International Joint Conference on Artificial Intelligence},
  pages={1923--1929},
  year={2022}
}

@inproceedings{cheng2025sourcedetmamba,
  author= {Le Cheng and
                  Peican Zhu and
                  Yangming Guo and
                  Chao Gao and
                  Zhen Wang and
                  Keke Tang},
  title        = {SourceDetMamba: {A} Graph-aware State Space Model for Source Detection
                  in Sequential Hypergraphs},
  booktitle    = {Proceedings of the 34th International Joint Conference on Artificial Intelligence},
  pages        = {2749--2757},
  year         = {2025}
}

@inproceedings{he2025hhan,
  title={Hhan: Comprehensive infectious disease source tracing via heterogeneous hypergraph neural network},
  author={He, Qiang and Bao, Yunting and Fang, Hui and Lin, Yuting and Sun, Hao},
  booktitle={Proceedings of the AAAI Conference on Artificial Intelligence},
  pages={291--299},
  year={2025}
}

@inproceedings{chen2021litegt,
  title={Litegt: Efficient and lightweight graph transformers},
  author={Chen, Cong and Tao, Chaofan and Wong, Ngai},
  booktitle={Proceedings of the 30th ACM International Conference on Information \& Knowledge Management},
  pages={161--170},
  year={2021}
}

@article{shelke2019source,
  title={Source detection of rumor in social network--a review},
  author={Shelke, Sushila and Attar, Vahida},
  journal={Online Social Networks and Media},
  volume={9},
  pages={30--42},
  year={2019},
  publisher={Elsevier}
}

@article{zoo_111,
  author       = {Forsyth, Richard},
  year         = {1990},
  howpublished = {UCI Machine Learning Repository}
}

@article{chodrow2021generative,
  title={Generative hypergraph clustering: From blockmodels to modularity},
  author={Chodrow, Philip S and Veldt, Nate and Benson, Austin R},
  journal={Science Advances},
  volume={7},
  number={28},
  pages={1303},
  year={2021},
  publisher={American Association for the Advancement of Science}
}

@article{chen2003visual,
  title={On visual similarity based 3D model retrieval},
  author={Chen, Ding-Yun and Tian, Xiao-Pei and Shen, Yu-Te and Ouhyoung, Ming},
  journal = {Computer Graphics Forum},
  volume={22},
  number={3},
  pages={223--232},
  year={2003},
}

@inproceedings{yadati2019hypergcn,
  title={Hypergcn: A new method for training graph convolutional networks on hypergraphs},
  author={Yadati, Naganand and Nimishakavi, Madhav and Yadav, Prateek and Nitin, Vikram and Louis, Anand and Talukdar, Partha},
  journal={Advances in Neural Information Processing Systems},
  pages={1509-1520},
  year={2019}
}

@article{mccallum2000automating,
  title={Automating the construction of internet portals with machine learning},
  author={McCallum, Andrew Kachites and Nigam, Kamal and Rennie, Jason and Seymore, Kristie},
  journal={Information Retrieval},
  volume={3},
  number={2},
  pages={127--163},
  year={2000},
  publisher={Springer}
}

@inproceedings{leskovec2012learning,
  title={Learning to discover social circles in ego networks},
  author={Leskovec, Jure and Mcauley, Julian},
  booktitle={Advances in Neural Information Processing Systems},
  pages={127--163},
  year={2012}
}

@inproceedings{cheng2025hyperdet,
author = {Cheng, Le and Zhu, Peican and Guo, Yangming and Tang, Keke and Gao, Chao and Wang, Zhen},
title = {HyperDet: source detection in hypergraphs via interactive relationship construction and feature-rich attention fusion},
year = {2025},
isbn = {978-1-956792-06-5},
booktitle = {Proceedings of the 34th International Joint Conference on Artificial Intelligence},
pages = {2758--2766}
}

@inproceedings{xia2022self,
  title={Self-supervised hypergraph transformer for recommender systems},
  author={Xia, Lianghao and Huang, Chao and Zhang, Chuxu},
  booktitle={Proceedings of the 30th ACM SIGKDD Conference on Knowledge Discovery and Data Mining},
  pages={2100--2109},
  year={2022}
}

@article{albert2000error,
	title={Error and attack tolerance of complex networks},
	author={Albert, R{\'e}ka and Jeong, Hawoong and Barab{\'a}si, Albert-L{\'a}szl{\'o}},
	journal={Nature},
	volume={406},
	number={6794},
	pages={378--382},
	year={2000},
	publisher={Nature Publishing Group UK London}
}

@inproceedings{louni2014two,
	title={A two-stage algorithm to estimate the source of information diffusion in social media networks},
	author={Louni, Alireza and Subbalakshmi, KP},
	booktitle={2014 IEEE Conference on Computer Communications Workshops},
	pages={329--333},
	year={2014},
}

@article{freeman2002centrality,
	title={Centrality in social networks: Conceptual clarification},
	author={Linton C. Freeman},
	journal={Social Networks},
	volume={1},
    number = {3},
	pages={215-239},
	year={2002}
}

@article{gajewski2022comparison,
	title={Comparison of observer based methods for source localisation in complex networks},
	author={Gajewski, {\L}ukasz G and Paluch, Robert and Suchecki, Krzysztof and Sulik, Adam and Szymanski, Boleslaw K and Ho{\l}yst, Janusz A},
	journal={Scientific Reports},
	volume={12},
	number={1},
	pages={5079},
	year={2022},
	publisher={Nature Publishing Group UK London}
}

@article{cheng2024sdsi,
  title={SDSI: Source Detection in Structurally Incomplete Social Networks},
  author={Cheng, Le and Zhu, Peican and Gao, Chao and Wang, Zhen and Li, Xuelong},
  journal={IEEE Transactions on Network Science and Engineering},
  year={2024},
  volume={12},
	number={2},
	pages={1041-1052},
  publisher={IEEE}
}

@article{hu2024source,
  title={Source localization in complex networks with optimal observers based on maximum entropy sampling},
  author={Hu, Zhao-Long and Wang, Hong-Jue and Sun, Lei and Tang, Chang-Bing and Li, Minglu},
  journal={Expert Systems with Applications},
  volume={256},
  number={124946},
  pages = {2918--2926},
  year={2024},
  publisher={Elsevier}
}

@inproceedings{DBLP:conf/nips/ZhouHS06,
   author = {Zhou, Dengyong and Huang, Jiayuan and Sch\"{o}lkopf, Bernhard},
 booktitle = {Advances in Neural Information Processing Systems},
 title = {Learning with Hypergraphs: Clustering, Classification, and Embedding},
 pages={378--382},
 year = {2006}
}

@inproceedings{feng2019hypergraph,
  title={Hypergraph neural networks},
  author={Feng, Yifan and You, Haoxuan and Zhang, Zizhao and Ji, Rongrong and Gao, Yue},
  booktitle={Proceedings of the AAAI conference on Artificial Intelligence},
  pages={3558--3565},
  year={2019}
}

@article{DBLP:journals/pr/BaiZT21,
    author       = {Song Bai and
                  Feihu Zhang and
                  Philip H. S. Torr},
    title = {Hypergraph convolution and hypergraph attention},
    journal = {Pattern Recognition},
    volume = {110},
    number = {107637},
    year = {2021},
}

@inproceedings{DBLP:conf/ijcai/HuangY21,
  author       = {Jing Huang and
                  Jie Yang},
  title        = {UniGNN: a Unified Framework for Graph and Hypergraph Neural Networks},
  booktitle    = {Proceedings of the 30th International Joint Conference on Artificial Intelligence},
  pages        = {2563--2569},
  year         = {2021}
}

@inproceedings{DBLP:conf/iclr/ChienPPM22,
  author       = {Eli Chien and
                  Chao Pan and
                  Jianhao Peng and
                  Olgica Milenkovic},
  title        = {You are AllSet: {A} Multiset Function Framework for Hypergraph Neural Networks},
  booktitle    = {Proceedings of the 10th International Conference on Learning Representations},
  year         = {2022},
  pages     = {2918--2926}
}

@inproceedings{DBLP:conf/cvpr/LiuCSBA22,
  author       = {Yuejiang Liu and
                  Riccardo Cadei and
                  Jonas Schweizer and
                  Sherwin Bahmani and
                  Alexandre Alahi},
  title        = {Towards Robust and Adaptive Motion Forecasting: {A} Causal Representation
                  Perspective},
  booktitle    = {Proceedings of the IEEE/CVF Conference on Computer Vision and Pattern Recognition},
  pages        = {17060--17071},
  year         = {2022}
}

@inproceedings{DBLP:conf/mm/0008CHYWY023,
  author    = {Shanshan Wang and Yiyang Chen and Zhenwei He and Xun Yang and
               Mengzhu Wang and Quanzeng You and Xingyi Zhang},
  title     = {Disentangled Representation Learning with Causality for Unsupervised
               Domain Adaptation},
  booktitle = {Proceedings of the 31st {ACM} International Conference on Multimedia
            },
  pages     = {2918--2926},
  year      = {2023}
}

@inproceedings{DBLP:conf/mm/Zhang00ZLH024,
  author       = {Litian Zhang and
                  Xiaoming Zhang and
                  Chaozhuo Li and
                  Ziyi Zhou and
                  Jiacheng Liu and
                  Feiran Huang and
                  Xi Zhang},
  title        = {Mitigating Social Hazards: Early Detection of Fake News via Diffusion-Guided
                  Propagation Path Generation},
  booktitle    = {Proceedings of the 32st {ACM} International Conference on Multimedia},
  pages        = {2842--2851},
  year         = {2024}
}

@inproceedings{DBLP:conf/mm/ZhaoNCB23,
  author       = {Wanqing Zhao and
                  Yuta Nakashima and
                  Haiyuan Chen and
                  Noboru Babaguchi},
  title        = {Enhancing Fake News Detection in Social Media via Label Propagation
                  on Cross-modal Tweet Graph},
  booktitle    = {Proceedings of the 31st {ACM} International Conference on Multimedia},
  pages        = {2400--2408},
  year         = {2023}
}

@inproceedings{DBLP:conf/mm/Wang0024,
  author       = {Yuzhuo Wang and
                  Junwei He and
                  Hongzhi Wang},
  title        = {{RHKH:} Relational Hypergraph Neural Network for Link Prediction on
                  N-ary Knowledge Hypergraph},
  booktitle    = {Proceedings of the 32st {ACM} International Conference on Multimedia},
  pages        = {8759--8767},
  year         = {2024},
}

@ARTICLE{11543370,
  author={Kang, Yan and Cheng, Le and Zhu, Peican and Liu, Zun and Long, Jiang and Guo, Yangming},
  journal={IEEE Transactions on Emerging Topics in Computational Intelligence}, 
  title={Enhanced Source Detection in Incomplete Networks via Redundancy-Guided Sensing and Anchor-Bounded Inference}, 
  year={2026},
  volume={},
  number={},
  pages={1-13}
}

@article{kang2026multi,
  title={Multi-source localization in social networks via hybrid sampling and Bayesian-inspired graph optimization},
  author={Kang, Yan and Zhu, Peican and Dong, Ang and Long, Jiang and Liu, Zun and Guo, Yangming},
  journal={Nonlinear Dynamics},
  volume={114},
  number={894},
  pages = {},
  year={2026}
}

\end{document}